\pdfoutput=1

\documentclass[aps,prb,twocolumn,showpacs,groupaddress]{revtex4}
\pdfoutput=1
\usepackage{amsmath}
\usepackage{amssymb}
\usepackage{epsfig}
\usepackage{bm}
\usepackage{longtable}
\usepackage{graphicx}
\usepackage{graphics}

\begin{document}

\title{
Exchange and correlation effects in the transmission phase through a 
few-electron quantum dot 
} 

\author{Massimo Rontani}
\email{rontani@unimore.it}
\affiliation{S3, Istituto di Nanoscienze -- CNR,
Via Campi 213/A, 41125 Modena, Italy}

\date{\today}

\begin{abstract}
The transmission phase through a
quantum dot with few electrons shows a complex, non-universal 
behavior. Here we combine configuration-interaction calculations 
---treating rigorously Coulomb interaction--- and   
the Friedel sum rule to provide a rationale for 
the experimental findings. 
The phase evolution for more than two electrons is found to  
strongly depend on dot's shape and electron density, 
whereas from one to two the phase never lapses.
In the Coulomb (Kondo) regime the phase shifts are
significant fractions of $\pi$ ($\pi/2$) for the second and subsequent charge 
addition if the dot is strongly correlated. These results 
are explained by the proper inclusion in the theory of Coulomb interaction,
spin, and orbital degrees of freedom.
\end{abstract}


\maketitle

\section{Introduction}

Recent experiments by the Weizmann group have stirred much attention
to the phase that electrons acquire when they
traverse a quantum dot (QD) embedded in the arm of a
Aharonov-Bohm interferometer.\cite{Moty, Zaffalon} These fascinating
measurements call for a deeper understanding of electron
transport through a strongly interacting object.
If many electrons --- say $N>10$ --- populate a QD, 
the transmission phase $\Theta$ of the 
tunneling electron displays a much studied\cite{Schuster97, Hackenbroich01}
universal behavior: first $\Theta$ increases by $\pi$ through the 
conductance peak and then it lapses in the Coulomb valley. 
Here we focus on the relatively less studied few-electron regime, 
where the phase evolution depends on $N$.
In the samples studied in Ref.~\onlinecite{Moty}
the phase remains constant in the $N=1$ valley,
independently from QD tunings,
whereas it lapses in the Coulomb blockaded devices
of Ref.~\onlinecite{Zaffalon}.
For $N>2$ the phase is not 
reproducible and shows 
both increments and lapses, likely due to QD shape 
and exchange effects.\cite{Moty}

In spite of several explanations of the few-electron 
scenario\cite{Moty, Karrasch07, Hecht09, Meden05, Golosov06, 
Entin-Wohlman00, Bertoni07, Silvestrov07, Gurvitz07, Yahalom06, Baksmaty08} 
a clear picture is still missing.
Many works introduced major simplifications, like spinless 
electrons,\cite{Karrasch07,Meden05,Golosov06} one-dimensional 
QDs,\cite{Entin-Wohlman00,Bertoni07} 
simplified models for Coulomb 
interaction,\cite{Karrasch07,Silvestrov07,Gurvitz07, Hecht09} or other
\emph{ad hoc} assumptions.\cite{Yahalom06} 
Even the interpretation of the simplest
$N=1\rightarrow N=2$ transition is controversial, being variably
attributed to the occupation of either the same\cite{Zaffalon} 
or a different\cite{Moty} orbital from
that of the first electron, 
to the role of excited doorway
channels,\cite{Baksmaty08} to electron crystallization.\cite{Gurvitz07}

In this paper we compute $\Theta$ by fully including  
exchange and correlation effects.
The theory is based on the application of the Friedel sum rule (FSR)
as generalized in Ref.~\onlinecite{RontaniFSR} to a multiorbital
interacting QD---an exact zero-temperature result, $T=0$. 
Since the FSR holds for non degenerate ground states (GSs)
only,\cite{RontaniFSR, Langreth66}  
it may be applied to either singlet Kondo GSs 
at zero field, $B=0$, or 
non-degenerate GSs in
the Coulomb blockade regime ($B\ge 0$ for singlets and $B>0$ for 
doublets and higher-spin states).

It is worth recalling that, in the conductance
valleys with odd $N$ at 
$T=0$ and $B=0$,
the QD is always in the Kondo regime. In order to reach
the Coulomb blockade regime by keeping $T=0$, one needs to apply 
the field to the Aharonov-Bohm ring to destroy dot-lead
Kondo correlations, hence
removing QD spin degeneracy.\cite{Ng88}
This condition is reached when  $\Gamma \ll \mu_B B $,
with $\mu_{\text{B}}$ being the Bohr magneton 
and $\Gamma$ the QD level width.
In the experiments the temperature is very low 
($T \sim 30$ mK) and the field is relatively weak ($B \sim 10$ mT).
Therefore, in order to compare measurements
with the theoretical predictions reported
in this paper, tiny widths $\Gamma$ are required,
which are actually smaller
than the values presently reported.\cite{Moty, Zaffalon}

As anticipated above, the theory relies on the application
of the FSR.
The key idea is that
the phase variation $\Delta\Theta$ for the addition 
of one electron to the QD is given by integrating the QD spectral density 
${\cal N}(\omega)$ between two consecutive
conductance valleys (with $\Delta N = 1$).
This is evaluated exactly for an isolated QD via full
configuration interaction (CI) calculations\cite{Rontani06} 
with $N\le 5$.

The simulations agree with the Coulomb-blockade results of
Ref.~\onlinecite{Moty}:
(i) The phase evolution for $2\le N\le 5$ strongly  
depends on the dot's shape and density.
(ii) $\Theta$ never lapses in the $N=1$ valley, independently from
dot parameters. Additionaly, (iii) in the Coulomb (Kondo) regime the
increment $\Delta \Theta$ through
the conductance peak is significantly smaller
than $\pi$ ($\pi/2$) as a consequence of strong Coulomb correlation.

The theory fails to reproduce the Coulomb
blockade results of Ref.~\onlinecite{Zaffalon}. The reason of this 
discrepancy is presently unclear, but it is likely related to
the significant departure from the condition
ruling the range of applicability of the theory, i.e.,
$\Gamma \ll \mu_B B $ at $T=0$.  
In order to treat the Coulomb blockade regime even at $B=0$,
one could compute the thermal Green function of the fully correlated system
at $T>T_K$, with $T_K$ being the Kondo temperature.
We leave for future work this alternative route,
which would immediately provide the transmission phase $\Theta$
without invoking the FSR.\cite{RontaniFSR}

The structure of this paper is as follows. In Sec.~\ref{s:FSR}
we set the theoretical model and illustrate the usage of the FSR
to compute the transmission phase variation $\Delta \Theta$
between consecutive conductance valleys. 
After briefly recalling the full CI method
in Sec.~\ref{s:CI}, we address in
Sec.~\ref{s:SP} the evolution of the phase in the absence
of Coulomb interaction (but taking into account the spin degree
of freedom). We eventually consider the fully interacting case
in Sec.~\ref{s:corr}.

\section{Transmission phase from the Friedel sum rule}\label{s:FSR}

\begin{figure}
\includegraphics[width=3.0in]{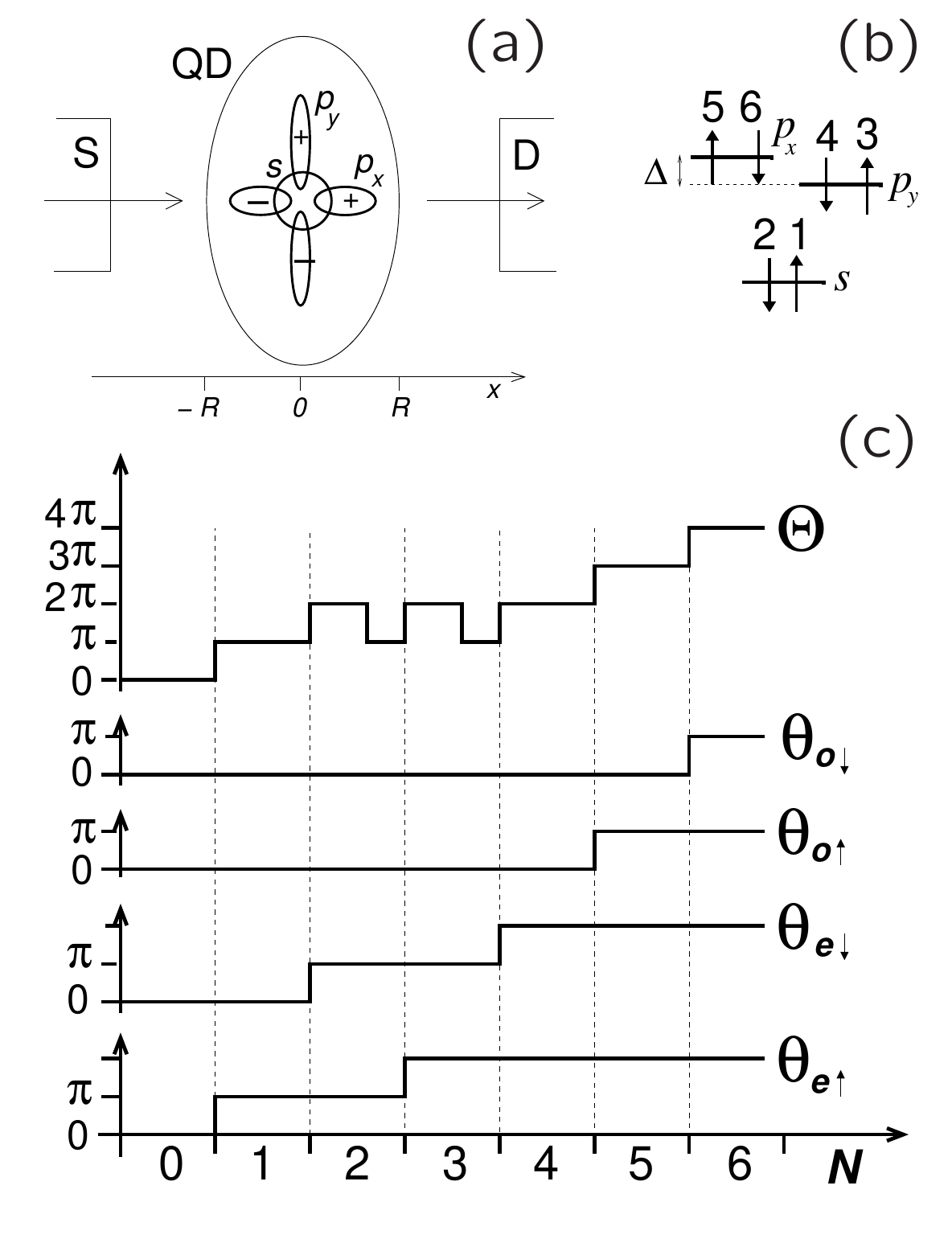}
\caption{(a) Experimental setup. 
(b) QD single-particle levels for $a/b<1$.
The number sequence points to the consecutive
filling of six electrons in a non-interacting picture.
(c) Transmission phase $\Theta$ and phase shifts
$\Delta \theta_{X\sigma}$ vs $N$ for the filling sequence 
plotted in (b).}
\label{Fig1}
\end{figure}

We model the experimental setup of Ref.~\onlinecite{Moty} 
as in Fig.~\ref{Fig1}(a).
Electrons flow along $x$ from source (S) to drain (D), tunneling
through a QD of elliptical shape. The ellipse is the generic 
low-energy form for a two-dimensional shallow, gate-defined 
potential,\cite{Moty, Zaffalon}
since the lowest-order non-vanishing terms of its series
expansion are quadratic.
The scattering matrix $S_{\sigma}$
is diagonal in the spin index $\sigma$ in both Kondo and Coulomb
blockade regimes.
In fact, in the Kondo regime ($B=0$),
no elastic spin-flip occurs.\cite{Langreth66,Ng88}
On the other hand, in the Coulomb blockade regime 
only one $\sigma$ channel is active at time for a given 
energy since $B$ removes spin degeneracy.

Additionaly, we assume\cite{NoteSymmetry} 
mirror reflection symmetry in the $yz$ plane
placed in the QD center, at $x=0$ [Fig.~\ref{Fig1}(a)],
hence the stationary scattering states $\psi_{\sigma}(x)$,
eigenstates of $S_{\sigma}$, are either 
even ($e$) or odd ($o$) with respect to $x\rightarrow -x$ 
reflection:
for $\left|x\right|> R$, 
\begin{displaymath}
\psi_{e\sigma}(x)=e^{-ik\left|x\right|}
+e^{2i\theta_{e\sigma}}e^{ik\left|x\right|} 
\end{displaymath}
and 
\begin{displaymath}
\psi_{o\sigma}(x)=\text{sgn}(x) \left[e^{-ik\left|x\right|}
+e^{2i\theta_{o\sigma}}e^{ik\left|x\right|}\right] .
\end{displaymath}
Here the even and odd
outgoing waves are phase shifted by $\theta_{e\sigma}$ and 
$\theta_{o\sigma}$, respectively, 
$k$ is the wave vector,
$R$ is the QD nominal longitudinal axis.
The left and right travelling states $\psi_{\pm k\sigma}(x)$
are superpositions
of even and odd states: 
\begin{displaymath}
\psi_{\pm k\sigma}(x)=\frac{1}{2}
\left[\psi_{e\sigma}(x)\mp\psi_{o\sigma}(x)\right].
\end{displaymath}
Inside the QD, $\left|x\right|< R$,
electrons experience  
two-body Coulomb interactions in addition to the confinement 
potential [cf.~Eq.~\eqref{eq:HI}].

We first generalize the results 
of Ref.~\onlinecite{Lee99} for spinless, non interacting electrons
by including spins.
The transmission amplitude $t_{\sigma}$ 
for travelling states 
$\psi_{\pm k\sigma}(x)$ is:
\begin{equation}
t_{\sigma}=\left|t_{\sigma}\right|ie^{i\Theta} 
=ie^{i\left(\theta_{e\sigma}+\theta_{o\sigma}\right)}
\sin{\left(\theta_{e\sigma}-\theta_{o\sigma}\right)}.
\label{eq:t}
\end{equation}
Each time $\sin{\left(\theta_{e\sigma}-\theta_{o\sigma}\right)}$
appearing in Eq.~(\ref{eq:t}) changes sign, due to a variation of either 
$\theta_{e\sigma}$ or $\theta_{o\sigma}$ as a new
electron tunnels into the QD, then a lapse
of $\pi$ occurs for the transmission phase $\Theta$. 
Since this happens when $t_{\sigma}=0$ [cf.~Eq.~\eqref{eq:t}], the
lapse is located in the conductance valley.\cite{Lee99, Taniguchi99}

We then include all many-body correlations by
connecting the phase shift $\theta_{X \sigma}$
per channel $(X,\sigma)$ ($X=e,o$ labels the parity) 
to the exact spectral density ${\mathcal{N}}_{X \sigma}(\omega)$
accumulated at the QD via the FSR [cf.~Eq.~(20) of 
Ref.~\onlinecite{RontaniFSR}]:
\begin{equation}
\frac{1}{\hbar}\frac{d\, \theta_{X \sigma}(\omega)}{d \omega}=
\pi{\mathcal{N}}_{X \sigma}(\omega).
\end{equation}
Here ${\mathcal{N}}_{X \sigma}(\omega)$ is the density
displaced at the QD by the electron in the scattering
state $(X,\sigma)$ tunneling at the 
energy $\hbar\omega$ fixed by the chemical potential 
$\mu= \hbar\omega$.
We mimic the action of the plunger gate
in the linear response regime by varying the value of $\mu$
with respect to the QD energy levels.

In practice, to use the FSR we integrate it
over the energy window between two consecutive Coulomb valleys
with $N$ and $N+1$ electrons in the QD, respectively.\cite{notaFSR}
Whereas this procedure provides the information on the
total phase variation only, 
$\Delta\theta_{X\sigma}=\pi\, \Delta {\mathcal{N}}_{X \sigma}=
\pi\hbar\int d\omega {\mathcal{N}}_{X \sigma}(\omega)$,
it allows to compute
$\Delta {\mathcal{N}}_{X \sigma}$ from the interacting Hamiltonian 
of the \emph{isolated} dot, 
$H_{\text{QD}}$. This key result is based on the conservation
of the total number of scattering plus QD states 
both in the presence and absence of the QD in the arm of the
interferometer.\cite{RontaniFSR} 

The CI evaluation of $\Delta \theta_{X \sigma}$ relies
on the formula
\begin{equation}
\Delta {\mathcal{N}}_{X \sigma} =
\frac{\Delta \theta_{X \sigma}}{\pi} = \sum_{\alpha_X}
\left|\left<\Psi_0^{N+1}|c^{\dagger}_{\alpha_X\sigma}
|\Psi_0^N\right>\right|^2,
\label{eq:exemplum}
\end{equation}
where $\left|\Psi_0^N\right>$ is the exact interacting 
GS of the isolated
QD with $N$ electrons of energy $E_0^N$, 
$H_{\text{QD}}\left|\Psi_0^N\right>=E_0^N
\left|\Psi_0^N\right>$, and $c^{\dagger}_{\alpha_X\sigma}$ creates 
an electron with spin $\sigma$ in the orbital of given 
parity $X$ and further specified by the set of quantum numbers $\alpha_X$.
Equation \eqref{eq:exemplum} follows from 
Eq.~(21) of Ref.~\onlinecite{RontaniFSR}, which was
inferred by connecting the phase shift
to the delay time spent by the electron wave packet in the QD. This delay
is obtained by integrating the wave function square modulus over both
time and space. By orthogonality of QD orbitals, only terms
diagonal in $\alpha_X$ indices survive in the formula \eqref{eq:exemplum}.

We eventually link 
the transmission phase variation 
$\Delta \Theta$ to $\Delta\theta_{X\sigma}$  
through Eq.~(\ref{eq:t}). 
In the Coulomb blockade regime only one scattering channel 
$(X,\sigma)$ is active at time between two consecutive valleys with 
respectively $N$ and $N+1$ electrons. The active channel is
univocally determined by the total spins and parities of 
$\left|\Psi_0^N\right>$ and $\left|\Psi_0^{N+1}\right>$,
as obtained by CI. On the other hand, in the Kondo regime time-reversal 
invariance (recall that $B=0$)
implies that $\Delta {\mathcal{N}}_{X \uparrow} = 
\Delta {\mathcal{N}}_{X \downarrow}$, evaluated as
half the Coulomb blockade value given by
Eq.~(\ref{eq:exemplum}). In this way we regain at
once the result\cite{Zaffalon, Langreth66} that
$\Delta \Theta = \pi/2$ for the addition of the first 
electron. In fact, the term on the right hand side of Eq.~(\ref{eq:exemplum})
is trivially one when $N=0$.

\section{The full configuration interaction method}\label{s:CI}

The interacting Hamiltonian of the isolated QD is
\begin{equation}
H_{\text{QD}} = \sum_{i=1}^{N}H_{\text{SP}}(i)
+\frac{1}{2}\sum_{i\neq j}\frac{e^{2}}
{\kappa|\bm{r}_i-\bm{r}_j|},
\label{eq:HI}
\end{equation}
where the single particle (SP) term is
\begin{equation}
H_{\text{SP}}(i)=\frac{ \bm{p}_i^2 }{ 2m^{*} } 
+ \frac{1}{2}m^{*}\!\left(\omega_{0x}^2x_i^2 + 
\omega_{0y}^2y_i^2\right) + 
\frac{1}{2} g^*\sigma_i\,\mu_BB.
\label{eq:HSP}
\end{equation}
Here $\sigma_i = \pm 1$, $\kappa$ is the dielectric constant,
$m^*$ is the electron effective mass, 
$g^*$ is the gyromagnetic factor,
and the QD confinement frequencies 
in the $x$ and $y$ directions, $\omega_{0x}$ and
$\omega_{0y}$, have characteristic lengths 
$a=[\hbar/(m^*\omega_{0x})]^{1/2}$
and  $b=[\hbar/(m^*\omega_{0y})]^{1/2}$, respectively
(the ratio $a/b$ is related to the ellipse eccentrity). 
In Eq.~(\ref{eq:HI}) the weak $B$ does not affect orbital 
degrees of freedom.

To wholly include in our theory Coulomb correlation,
we solve numerically the few-body problem of Eq.~(\ref{eq:HI})
by means of the full CI method
(also known as exact diagonalization,
for details see Ref.~\onlinecite{Rontani06}).
The CI few-body GS $\left|\Psi_0^N\right>$ is essentially
a linear combination of the Slater determinants
$\left|\Phi_i^N\right>$,
\begin{equation}
\left|\Psi_0^N\right> = \sum_i c_i \left|\Phi_i^N\right>,
\label{eq:CIexp}
\end{equation}
with the unknown $c_i$s being the output of the calculation.
Here the determinants $\left|\Phi_i^N\right>$ are
obtained by filling in all possible ways with $N$ electrons
the $N_{\text{SP}}$ lowest-energy SP orbitals (two-fold
spin degenerate at $B=0$), eigenstates of the
SP Hamiltonian \eqref{eq:HSP}. In the Fock space of these
Slater determinants $H_{\text{QD}}$
is a large sparse matrix, that we exactly diagonalize
by means of the parallel code \emph{DonRodrigo},\cite{website}
eventually obtaining the coefficients $c_i$ of Eq.~\eqref{eq:CIexp}.

The diagonalization proceeds in each Hilbert space sector 
labeled by $N$, the total spin, and the total parity of the
few-body wave function. After we have obtained the GSs
$\left|\Psi_0^N\right>$ and $\left|\Psi_0^{N+1}\right>$,
we evaluate $\Delta {\mathcal{N}}_{X \sigma}$
via Eq.~(\ref{eq:exemplum}), and
eventually infer $\Delta \Theta$ as explained in Sec.~\ref{s:FSR}.

In the CI calculations reported in Sec.~\ref{s:corr}
we used $N_{\text{SP}}=36$ and diagonalized matrices
of maximum linear size 2.25 $\times 10^6$.
The relative error for the energy was less
than $10^{-4}$ for $a/b=1$.

\section{The spinful non-interacting case}\label{s:SP}

To illustrate the effect of the inclusion of the spin degree of freedom
in the calculation of the transmission phase
$\Theta$ let us consider for the time being
only the SP Hamiltonian, $H_{\text{SP}}$,
and neglect Coulomb interaction.
The GS is a Slater determinant 
with the lowest $N$ spin-orbitals filled, 
$\left|\Psi^N_0\right>=\prod_{i=1}^Nc^{\dagger}_{\alpha_i\sigma_i}\!
\left|0\right>$ ($\left|0\right>$ is the vacuum). 
The \emph{Aufbau} filling sequence for $1\le N\le 6$ is depicted in 
Fig.~\ref{Fig1}(b) for $a/b < 1$. The first two electrons occupy the
$s$ orbital with opposite spin, then
the 3rd and 4th electrons fill in the $p_y$ orbital, which is shifted
in energy from the $p_x$ orbital by $\Delta = \hbar\omega_{0x}
(1-a^2/b^2)$. Note that the first electron entering
a new SP level is always $\uparrow$, due to the effect of $B$. 
The evaluation of the phase shift $\Delta\theta_{X\sigma}$
at each electron addition is straightforward, since in 
Eq.~(\ref{eq:exemplum}) only one addendum gives a 
non-zero contribution to $\Delta\theta_{X\sigma}/\pi$ ---exactly one---
as a new spin-orbital $(\alpha_X,\sigma)$ is occupied;
the other ones vanish due to the orthogonality
of the states. Therefore in Fig.~\ref{Fig1}(b) one has
$ \Delta {\mathcal{N}}_{X \sigma}=1$ 
for the sequence $(X, \sigma) = ( e,\uparrow)$,
$(e,\downarrow)$, $(e,\uparrow)$, $(e,\downarrow)$,
$(o,\uparrow)$, $(o,\downarrow)$ of six consecutive electron
additions, with the $p_y$ ($p_x$) orbital even
(odd) under $x\rightarrow -x$.

The evolution of $\Theta$ for the 
filling sequence of Fig.~\ref{Fig1}(b)
is shown in Fig.~\ref{Fig1}(c).
Both increments and lapses of $\Theta$ are derived through  Eq.~(\ref{eq:t})
(lapse locations in the conductance valleys with fixed $N$ are arbitrary).
A remarkable feature of Fig.~\ref{Fig1}(c) is that 
$\Theta$ increases by $\pi$ in both transitions $N=0\rightarrow N =1$
and $N=1\rightarrow N =2$, since the first
two electrons occupy the same $s$ orbital with opposite spin.
This is fundamentally different from the spinless
case,\cite{Lee99} where a total increase of $\Theta$ by $2\pi$ 
between $N=0$ and $N=2$ occurs
only if the two electrons occupy orbitals of different parities.
 
Two lapses of $\pi$ occur for $\Theta$ in the blockaded regions 
with $N = 2$ and $N=3$ 
[Fig.~\ref{Fig1}(c)] as the phases $\theta_{e\uparrow}$
and $\theta_{e\downarrow}$ increase more than $\pi$, respectively.  
The whole pattern of $\Theta$ in Fig.~\ref{Fig1}(c) 
up to five electrons coincides with Fig.~4a
of Ref.~\onlinecite{Moty}, provided that 
one interprets the $N=3\rightarrow N=4$ smooth transition 
as a phase lapse (in Sec.~\ref{s:corr} we consider
an alternative interpretation). This agreement is surprising, since
in the experiment SP levels have
a small energy separation ($\sim 0.5$ meV), if compared to characteristic
Coulomb energies ($\sim 1-3$ meV),\cite{Moty}
and therefore one would expect significant differences
from the non-interacting model of Fig.~\ref{Fig1}.
On the other hand, the $\Theta$-evolution would be 
basically the same as in Fig.~\ref{Fig1} if
the interacting QD ground state were well approximated
by a single Slater determinant, as in Hartee-Fock
theory where Coulomb interaction is included as a mean field.

\section{The role of Coulomb interaction}\label{s:corr}

When correlation effects beyond the mean-field level\cite{explanation} 
are relevant,
we expect that $\Delta {\mathcal{N}}_{X \sigma} < 1$, as
suggested in Ref.~\onlinecite{RontaniFSR}.
Since the CI ground states $\left|\Psi_0^N\right>$ 
are superpositions of the Slater determinants  $\left|\Phi_i^N\right>$ 
[cf.~Eq.~\eqref{eq:CIexp}],
after expansion on this basis many cross terms give no contribution 
to Eq.~(\ref{eq:exemplum}):
the stronger the correlation, the larger the number of Slater 
determinants, the smaller  $\Delta {\mathcal{N}}_{X \sigma}$.
This seems to be the case in Ref.~\onlinecite{Moty}
for $\Delta \Theta \sim 3\pi/4$ in the $N=1\rightarrow N=2$ 
transition of Fig.~4b and $\Delta \Theta \sim 3\pi/4$
for $N = 6 \rightarrow N = 7$ in Fig. 5.
This could even be the case for
$\Delta \Theta \sim 0 $ for
$N = 3 \rightarrow N = 4$ in Fig.~4a of
Ref.~\onlinecite{Moty}, if one excludes
the possibility of a phase lapse. Such interpretation
is alternative to the one suggested 
in the previous section. 

\begin{figure}
\includegraphics[width=3.5in]{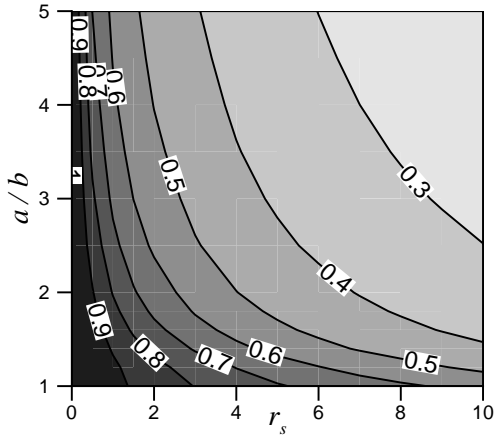}
\caption{Contour plot of $\Delta \Theta$ in the  $(r_s,\,a/b)$ 
space, in units of $\pi$, for the $N=1\rightarrow N=2$ transition
in the Coulomb blockade regime. The grey code goes from 0 (white) 
to 1 (black). The value in the Kondo regime is obtained
by dividing $\Delta \Theta$ by two.}
\label{FigContour}
\end{figure}

\begin{figure}
\includegraphics[width=3.0in]{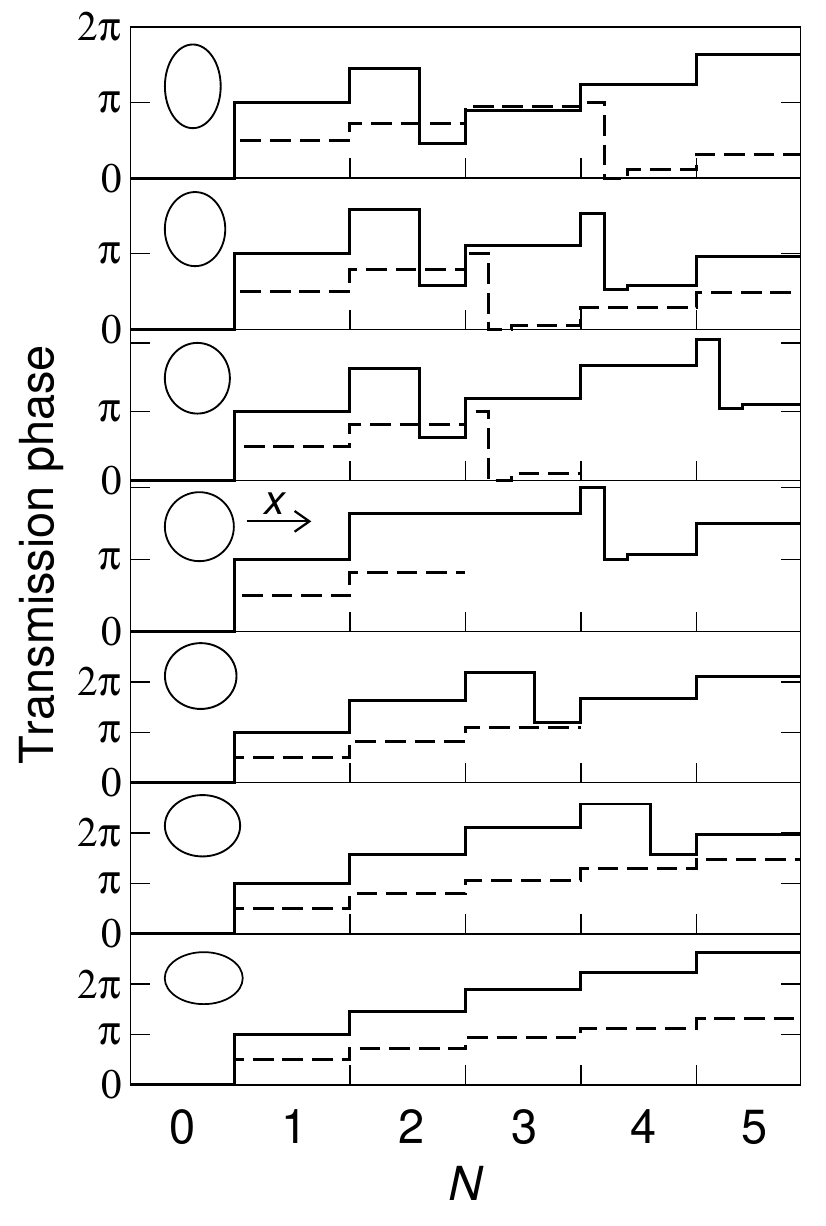}
\caption{$\Theta$ vs $N$ for different dot anisotropies
$a/b$. Solid (dashed) lines refer to the Coulomb blockade 
(Kondo) regime.
We used GaAs parameters and $\hbar\omega_0=0.5$ meV.
From top to bottom: $a/b=$ (1.5)$^{-1}$, (1.15)$^{-1}$, 
(1.05)$^{-1}$, 1, 1.05, 1.15, 1.5.}
\label{figLast}
\end{figure}

To assess the impact of correlation in the CI results,
we parameterize the electron density $n$ of the circular dot 
($\omega_0=\omega_{0x} =\omega_{0y}$) via
the dimensionless radius $r_s$ of the circle whose area is
equal to the area per electron, 
$r_s = 1/[a_{\text{B}}^*(\pi n)^{1/2}]$,
where $a_{\text{B}}^*$ is the effective Bohr radius and 
$n$ is estimated as in Ref.~\onlinecite{Garcia05}.
We next focus on the evolution of
$\Theta$ as a function of 
both $r_s$ and ellipse anisotropy ratio $a/b$.
The first electron addition in the Coulomb
(Kondo) regime always gives $\Delta \Theta = \pi$
($\Delta \Theta = \pi/2$). Then $\Theta$ remains constant 
in the $N=1$ valley, independently from the values 
of either $r_s$ or $a/b$ [cf.~Fig.~\ref{figLast}].
The second electron addition is analyzed in Fig. \ref{FigContour},
plotting in the $(r_s,\,a/b)$ space the contour map
of $\Delta \Theta$ for $N=1\rightarrow N=2$.
Here we vary the anisotropy ratio $a/b$ 
by keeping the ellipse area $\pi ab$ fixed, so the density remains constant.
The contour lines of Fig.~\ref{FigContour} provide
the value of $\Delta \Theta$ in the Coulomb blockade regime,
whereas its Kondo counterpart may be simply obtained by 
dividing $\Delta \Theta$ by two (cf.~also Fig. \ref{figLast}).
As $r_s$ increases, $\Delta \Theta$ monotonously decreases,
since correlation effects become stronger at lower density,
as the Coulomb term in $H_{\text{QD}}$
[Eq.~(\ref{eq:HI})] overcomes the SP term.
A similar trend occurs by increasing $a/b$, since a stronger
anisotropy effectively lowers the dimensionality of the system, 
again enforcing correlation effects.\cite{notaCN}
Note that by overlapping the experimental value $\Delta \Theta \sim 3\pi/4$
with the plot of Fig.~\ref{FigContour} we find $r_s\approx 4$ for $a/b=1$.
This value corresponds to $\hbar\omega_0=1.2$ meV for GaAs,
which is comparable to the experimental estimate of 0.5 meV.\cite{Moty}  

From the analysis of CI data of Fig.~\ref{FigContour}
we find that the orbital parities of both the first and second 
electrons are always even,
independently from $r_s$ and $a/b$.
This prediction agrees with the Wigner-Mattis theorem:
the two-electron GS is always
a singlet and the orbital part of its wave function is 
nodeless.\cite{Mattis65} This result conflicts with other 
explanations,\cite{Moty,Baksmaty08} and it is expected to hold even in the
presence of disorder and/or more complicated potentials. 

In Fig.~\ref{figLast} we follow the evolution of $\Theta$ up to
five electrons for a significant range of 
QD anisotropies. At the experimental density
($\hbar\omega_0=0.5$ meV) a slight variation of $a/b$
is sufficient to alter the phase behavior
for $N>2$.
Indeed, the relative differences between the values
of $a/b$ for the 3rd, 4th, and 5th panels are as small as 5\%.   
Therefore, $\Theta$ is sensitive
to fluctuations of the experimental QD parameters, as reported
in Ref.~\onlinecite{Moty}.

The occurrence of alternative scenarios in Fig.~\ref{figLast} 
is another signature of correlation.
In fact, several excited states lie very close in energy to the GS, 
as it is the case in the crossover to electron 
crystallization.\cite{Reimann02, Kalliakos08, Singha10} 
Hence a small deformation of the QD shape easily induces a crossing
between states of different symmetry.  
We here highlight only the most relevant features of a rich zoology,
focusing on Coulomb blockade results (solid lines in Fig.~\ref{figLast}).
In a circular QD at such low density (4th panel of Fig.~\ref{figLast}) 
the three-electron GS is a spin quadruplet as an effect of
correlation.\cite{Rontani06} Because the two-electron GS is a singlet,  
the transition $N=2\rightarrow N=3$ is spin blockaded,
i.e., $\Delta \Theta = 0$ without any lapse as
$\Delta {\mathcal{N}}_{X\sigma}=0$. A slight deformation of
the QD (3rd and 5th panels) changes the $N=3$ GS into a doublet,
lifting the spin blockade ($\Delta \Theta\neq 0$ between $N=2$
and $N=3$). 
The $N=4$ GS is a more robust triplet,
since the spin polarization is due to Hund's rule 
---an open shell effect.\cite{Reimann02}
However, a stronger deformation of the QD (2nd and 6th panels) breaks 
the orbital degeneracy of the SP levels of the 2nd shell 
inducing a transition
to a singlet GS. At such anisotropy ratios 
singlets and doublets typically alternate for even and odd electron
numbers, respectively. A further increase of 
the deformation (1st and 7th panels)
changes the filling sequence of higher-energy orbitals.

In Fig.~\ref{figLast} we also plot $\Theta$ in the
Kondo regime (dashed lines) for those cases 
such that the QD spin is totally screened by the cloud of
opposite-spin tunneling electrons.\cite{Hewson} 
This excludes high-spin GSs other than singlets
an doublets occuring   
in the 3rd, 4th, and 5th panels.
The hallmark of correlation is that 
$\Delta \Theta$ is a fraction of $\pi$ and $\pi/2$ in the
Coulomb blockade and Kondo regimes, respectively 
[e.g., compare the SP phase evolution of 
Fig.~\ref{Fig1}(c) with its correlated counterpart in
the 2nd panel of Fig.~\ref{figLast}].

\section{Conclusion}

In conclusion, we highlighted the role of exchange
and correlation in the transmission phase of a
few-electron quantum dot. Our findings are relevant 
for transport experiments through
strongly interacting nano objects, including 
molecules and carbon-based nanostructures.

\begin{acknowledgments}
We thank M. Heiblum, A. Bertoni, and A. Calzolari  
for discussions and L. Neri for proofreading the paper. 
This work was supported by INFM-CINECA 2008-2009.
\end{acknowledgments}


\begin{thebibliography}{35}
\bibitem{explanation} Here we disregard the relaxation of self-consistent
orbitals at the Hartree-Fock level, which may decrease
$\Delta {\mathcal{N}}_{X \sigma}$ from its unit value.
On the other hand, our CI results for 
$\Delta {\mathcal{N}}_{X \sigma}$ are exact, including both the
mean-field relaxation as well as the 
effect of correlations beyond mean-field. The latter are 
dominant at large values of $r_s$.
\end{thebibliography}
\end{document}